\documentclass{article}

\usepackage{microtype}
\usepackage{graphicx}
\usepackage{subfigure}
\usepackage{booktabs} 
\usepackage{amsmath}
\usepackage{svg}
\DeclareMathOperator*{\argmax}{\arg\!\max}

\usepackage{hyperref}
\usepackage[font=small,skip=0pt]{caption}

\usepackage[accepted]{icml2021}

\begin{document}

\twocolumn[
\icmltitle{Symmetry-Constrained Generation of Diverse Low-Bandgap Molecules with Monte Carlo Tree Search}



\icmlsetsymbol{equal}{*}

\begin{icmlauthorlist}
\icmlauthor{Akshay Subramanian}{dmse}
\icmlauthor{James Damewood}{dmse}
\icmlauthor{Juno Nam}{dmse}
\icmlauthor{Kevin P. Greenman}{cheme}
\icmlauthor{Avni P. Singhal}{dmse}
\icmlauthor{Rafael G{\'o}mez-Bombarelli}{dmse}
\end{icmlauthorlist}

\icmlaffiliation{dmse}{Department of Materials Science and Engineering, Massachusetts Institute of Technology, Cambridge, MA, USA}
\icmlaffiliation{cheme}{Department of Chemical Engineering, Massachussets Institute of Technology, Cambridge, MA, USA}

\icmlcorrespondingauthor{Rafael Gomez-Bombarelli}{rafagb@mit.edu}

\icmlkeywords{Organic Electronics, Machine Learning, Molecular Design, Atomistic Simulations}

\vskip 0.3in
]



\printAffiliationsAndNotice{}  

\begin{abstract}
Organic optoelectronic materials are a promising avenue for next-generation electronic devices due to their solution processability, mechanical flexibility, and tunable electronic properties. In particular, near-infrared (NIR) sensitive molecules have unique applications in night-vision equipment and biomedical imaging. Molecular engineering has played a crucial role in developing non-fullerene acceptors (NFAs) such as the Y-series molecules, which have significantly improved the power conversion efficiency (PCE) of solar cells and enhanced spectral coverage in the NIR region. However, systematically designing molecules with targeted optoelectronic properties while ensuring synthetic accessibility remains a challenge. To address this, we leverage structural priors from domain-focused, patent-mined datasets of organic electronic molecules using a symmetry-aware fragment decomposition algorithm and a fragment-constrained Monte Carlo Tree Search (MCTS) generator. Our approach generates candidates that retain symmetry constraints from the patent dataset, while also exhibiting red-shifted absorption, as validated by TD-DFT calculations.

\end{abstract}

\section{Introduction}

Organic optoelectronic materials have emerged as promising candidates for next-generation electronic devices, particularly in the fields of organic photodiodes (OPDs) and organic photovoltaics (OPVs). These materials offer unique advantages such as solution processability, mechanical flexibility, and tunable electronic properties, making them attractive alternatives to traditional inorganic semiconductors \cite{sun2022recent,li2022recent,shan2022organic}. Of particular interest are low-bandgap $\pi$-conjugated organic molecules and polymers sensitive to near-infrared (NIR) wavelengths, which can harvest a broader portion of the solar spectrum, extending beyond the visible range. These NIR-sensitive materials have found diverse applications, from military night-vision equipment to biomedical imaging devices and semi-transparent photovoltaics  \cite{Xu2008, Liu2017, Li2017, Xu2017}.
The optimization of molecular structures, especially the design of improved donor--acceptor (D-A) systems, has led to substantial progress in the power conversion efficiency (PCE) of OPV devices. Over the past decade, PCE has increased from around 5\% to over 19\% for single-junction devices \cite{Meng2022}. This advancement is largely attributed to the development of non-fullerene acceptors (NFAs), which offer more tunability of energy levels and absorption spectra through molecular engineering, than fullerene acceptors (FAs) \cite{yan2018non}. The Y-series family of NFAs (Y1 through Y6) has been a particularly promising class of molecules \cite{yang2021original}, which has sparked several follow-up designs and derivatives \cite{he2022molecular}. These improvements highlight the role of molecular design in enhancing device performance and the potential for further advancements through innovative approaches to materials discovery.

Over the past half-decade, generative models have received tremendous attention in molecular design \cite{subramanian2024closing,schwalbe2020generative,elton2019deep}, using a variety of algorithms and molecular representations \cite{segler2018generating,bjerrum2017molecular,gomez2018automatic,kusner2017grammar,olivecrona2017molecular, mercado2021graph,jin2018junction,flam2021mpgvae}. However, without special attention to synthetic accessibility, off-the-shelf models often generate non-synthesizable designs \cite{gao2020synthesizability}. Several existing approaches incorporate knowledge about synthesizability into generative models, either implicitly or explicitly.

Implicit approaches typically work by 1) biasing the objective function for property optimization with an additional synthesizability term, 2) applying a synthesizability filter post-hoc to generated candidates, or 3) constraining the training dataset using combinatorial enumerations of molecular fragments (for example, symmetric attachments as a proxy for synthetic accessibility). 1 and 2 often 
use approximate score functions such as SAscore \cite{ertl2009estimation}, SCScore \cite{coley2018scscore}, or retrosynthesis tools like ASKCOS \cite{coley2019robotic} or AiZynthFinder \cite{genheden2020aizynthfinder} to quantify synthesizability. In 1, multi-objective optimization is sensitive to the choice of the loss function's functional form, which if not done appropriately can lead to a compromise in other property objectives. In 2, a key assumption is that a sufficiently large portion of generated candidates are synthesizable, an assumption often proven false, as demonstrated in Figure S10 of \citet{gao2020synthesizability}. Additionally, the scoring functions themselves have specific definitions that make them more suitable for certain situations than others, limiting their universality. SAscore for instance is designed for biologically relevant molecules such as drug-like molecules, and might be less applicable for other chemical spaces such as organic electronics \cite{ertl2009estimation}. In 3, a downside is that the heuristics used to design the training dataset might not be reflected in the generations unless more explicit constraints are injected into the models.

Explicit methods of incorporating synthesizability on the other hand, do not suffer from limitations of scoring functions, or generations not satisfying prior constraints. They typically either embed pre-specified synthetic reaction templates \cite{gao2021amortized,hartenfeller2012dogs,vinkers2003synopsis,swanson2024generative}, or machine-learned reaction prediction algorithms \cite{bradshaw2019model,bradshaw2020barking} into the generation grammar to constrain the scope of generation. While both reduce the possibility of chemically infeasible outcomes, reaction templates significantly limit the breadth of the search space, and reaction prediction algorithms could be prone to error propagation in multi-step synthetic routes \cite{gao2020synthesizability}. This case of multi-step pathways is  common in organic electronics systems \cite{strieth2024delocalized}.
Some recent works at the intersection of organic electronics and synthesizability include  \citet{kwak2022design,li2023generative,westermayr2023high}.

In this paper, we develop a simple, and more universally applicable approach to incorporate an approximate form of synthesizability into the generation process, and demonstrate its utility for organic electronics applications by generating low-bandgap molecules. Building on our previous work on extracting datasets from published patents, our method is based on the fact that molecules reported in published patents are likely to be synthetically accessible, and when extracted based on domain-focused keyword queries made on the text, are also high-performance for domain-relevant properties \cite{subramanian2023automated}. The chemical fragments that make up a molecule, and the symmetry in their attachments contain key information about the precursors and synthetic pathways used to make the molecule. To utilize this information effectively during generation, we develop:
\begin{enumerate}
  \item A fragment decomposition algorithm that extracts fragments and their reactive positions from patent-mined structures, preserving symmetry information.
  \item A generation algorithm based on Monte Carlo Tree Search (MCTS) \cite{kocsis2006bandit,coulom2006efficient,swiechowski2023monte} that constrains the generated molecules to contain only the previously extracted fragments attached at their reactive positions, while satisfying symmetry requirements.
\end{enumerate}
We evaluate our method on two chemical spaces of different sizes, a smaller, focused space of Y6-derivatives \cite{yang2021original}, and a much larger space of patent-mined fragment derivatives \cite{subramanian2023automated}. We employ Time-Dependent Density Functional Theory (TD-DFT) to obtain our ``ground-truth'' bandgaps for training and evaluation. Our generated candidates retain symmetry constraints from the patent dataset while also exhibiting red-shifted absorption.

\section{Methods}
\subsection{Fragment Identity, Connectivity and Symmetry}
Our previous work developed an automated pipeline to mine USPTO patents and create domain-focused unsupervised datasets of molecular structures \cite{subramanian2023automated}. These patent-derived molecules likely possess two important characteristics: high performance (justifying the investment in a patent application) and synthetic accessibility. Consequently, the fragments present in these structures, along with their reactive positions (points of connectivity), potentially correlate with both synthetic accessibility and domain-specific properties. Moreover, the symmetry of reactive positions relative to each other encodes information about the synthetic pathways used to create these molecules.
To effectively leverage this structural and synthetic information, we have developed a two-stage method:
\textbf{Fragment Decomposition:} We designed an algorithm that extracts fragments and their reactive positions from patent-mined structures, while marking each reactive position in a way that faithfully represents the identity of the substructure attached to it. This preserves important contextual information about molecular connectivity. It is this connectivity and identity-preserving property that we refer to as "symmetry" throughout this paper.
\textbf{Constrained Generation:} We implemented a generation algorithm based on MCTS that constrains the output to molecules composed solely of patent-extracted fragments. These fragments are attached at their corresponding reactive positions while adhering to the symmetry requirements observed in the original patent structures. This approach aims to maintain synthetic feasibility while exploring novel molecular configurations.
By combining these two components, we aim to create a framework for generating molecules that are not only likely to be synthetically accessible but also tailored to the specific property requirements of organic electronic materials. This method leverages the information embedded in patent-derived structures to guide the exploration of chemical space towards promising and realizable candidates.

\begin{figure}[ht]
  \centering
  \includegraphics[width=0.45\textwidth]{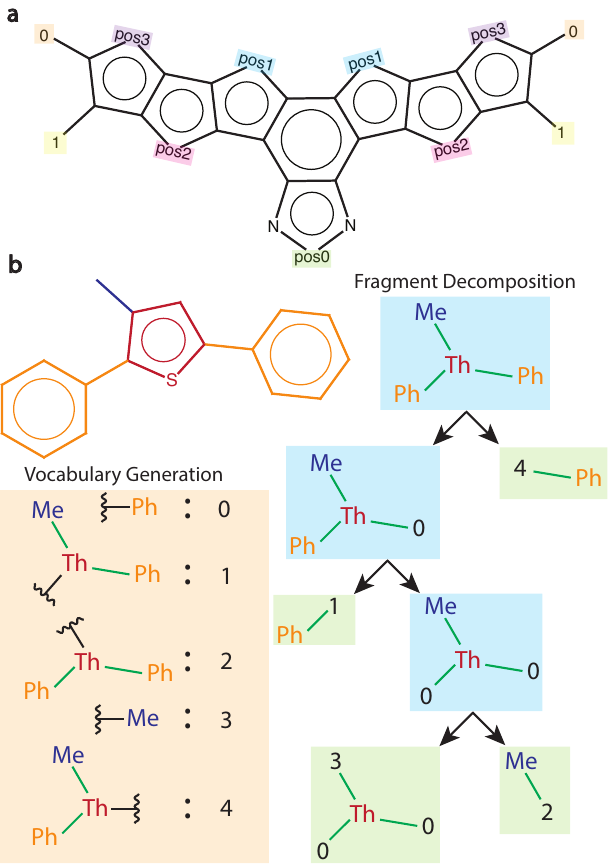}
  \caption{\textbf{Fragments to initialize MCTS} (a) Editable Y6 core with marked positions. (b) Fragment decomposition algorithm to obtain fragments from patent dataset. Vocabulary dictionary is first created by breaking one-bond at a time, and labeling the resulting fragments with unique integer values. Recursive decomposition of the starting molecule is then performed. Reactive positions are labeled with values corresponding to the broken fragment's identity. The leaf nodes (shaded in green) represent the final set of fragments obtained after decomposition. Me, Th, and Ph represent methyl, thiophene, and phenyl groups, respectively.  }
  \label{fig:frag_decomp}
\end{figure}

\subsubsection{Fragment Decomposition}
\label{decomp}

We developed a fragment decomposition algorithm that preserves connectivity and symmetry information from patent molecules. We achieve this through a recursive application of RDKit reaction SMARTS \cite{landrum2013rdkit}. More details are provided in the following paragraphs, and an illustrative example is shown in Figure \ref{fig:frag_decomp}(b).

A reaction SMARTS string is first defined to indicate what types of bonds must be deleted during decomposition. For our purpose, we defined this so that single bonds connecting an aromatic ring with an aliphatic chain, or an aromatic ring to another aromatic ring are matched.
The algorithm consists of two steps: 1) creation of fragment vocabulary, 2) fragment decomposition and assignment of reactive sites. 

In step 1, the reaction SMARTS is applied to the molecule to get all possible fragments that are obtainable from deletion of one bond in the molecule. The vocabulary of unique fragments is created, and each is assigned a unique ID. For the IDs, we use numbers $[0,N)$ where $N$ is the number of fragments in the vocabulary. Details on how this is actually represented in our code implementation are given in SI Section \ref{pos_practice}.

In step 2, the original molecule is recursively decomposed with one-bond deletions to form a binary tree. After every application of the reaction SMARTS, the reactive positions (positions at which bond deletion was performed) on the fragment are labeled using the ID of the fragment which was attached before the bond was broken. This preserves information about the identity of the fragment which was connected at the reactive position. The two fragments are then checked for a match with the reaction SMARTS to see if further decomposition is possible. If possible, the same series of steps described above is performed on the fragment. Otherwise, the fragment is a leaf node on the binary tree, and is added to the list of final fragments.

Finally, memoization \cite{michie1968memo} can be used to prevent redundant decompositions of the same fragment. We memoize computations in a hash table during tree creation, and reuse previously performed computations. Figure \ref{fig:frag_decomp} illustrates this process of fragment decomposition on an example molecule.

\subsubsection{Generation with MCTS}
We frame the problem of designing a molecule as a series of building block/molecular fragment choices. MCTS is an efficient method of navigating large exploration spaces, that learns to optimize a chosen reward function by iteratively updating value estimates for action choices in the tree \cite{kocsis2006bandit,coulom2006efficient,swiechowski2023monte,silver2016mastering}. Since our objective is to design low-bandgap molecules, our reward function of choice is the negative bandgap as predicted by Chemprop \cite{heid2023chemprop}, a message-passing neural network (MPNN) architecture that we trained on labels obtained from TD-DFT calculations. While more details about TD-DFT calculations are provided in Section \ref{tddft}, we note here that our use of the term "bandgap" throughout this paper refers to the lowest-energy singlet vertical excitation energies.
Each iteration of the MCTS algorithm consists of four sequential steps: selection, expansion, rollout, and backpropagation. 
The \textbf{selection step} starts with the root node and traverses the already explored portions of the tree to reach a leaf node. In our work, the path from the root node to the leaf node represents the construction of a molecular structure, where each node corresponds to a partial or complete structure. We follow the standard Upper Confidence Bound applied to Trees (UCT) policy \cite{kocsis2006bandit} to choose an action at each timestep of traversal. 
\begin{equation*}
\label{eq:UCT}
a_t = \argmax_{a \in A_t}\left[Q_{t-1}(a) + c\sqrt{\frac{\log(N^p_{t-1})}{N_{t-1}(a)}}\right]
\end{equation*}

where $a_t$ is the action at timestep $t$, $A_t$ is the set of action choices available at timestep $t$, $Q_{t-1}(a)$ is the action value function of action $a$ before timestep $t$, $N_{t-1}(a)$ is the number of times the action $a$ has been chosen before timestep $t$, and $N^p_{t-1}$ is the number of times the parent node has been visited before timestep $t$. The exploration coefficient $c$ controls the exploration-exploitation trade-off, i.e., the larger its value, the higher the exploration.

If the leaf node is not a terminal action, the \textbf{expansion step} enumerates possible next states and expands the tree with new nodes, one of which is then chosen at random.
In the \textbf{rollout step}, a random policy is followed until the molecular generation process is terminated.
Finally, the quality of the end state is evaluated and is assigned a scalar reward value. The reward is \textbf{backpropagated} through the tree branch to update the value functions of each visited node. 

A detailed description of the Markov Decision Process defined for MCTS is given in Section \ref{mdp}. The Y6 case does not involve patent-mined fragments, and hence symmetry and reactive position considerations are greatly simplified. However, in the patent-mined fragments case, these considerations need to be incorporated into the generation grammar. Only fragments obtained from the fragment decomposition algorithm (described in Section \ref{decomp}) are utilized in the generation process, and fragment attachments are made only to reactive position markers. Identically marked positions are attached with fragments of identical identities, while differently marked positions are attached with fragments of different identities, thereby preserving the connection symmetry as was present in the patent structure.



\begin{figure*}[ht]
  \centering
  \includegraphics[width=\textwidth]{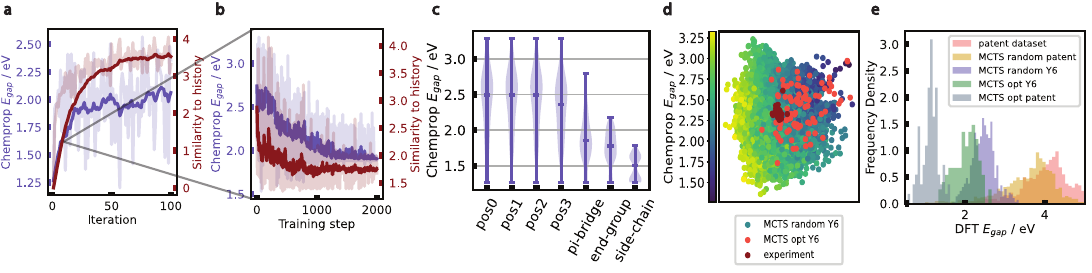}
  \caption{\textbf{Features of MCTS training and DFT validation} (a) Bandgap and similarity penalty as a function of iteration. Values shown are the weighted contributions to the reward function. With every iteration, the constrained optimization becomes more challenging resulting in less optimal candidates. (b) Representative reward evolution during training for one MCTS repetition. (c) Pruning of tree during training. The tree is expanded to contain deeper choices such as end-groups and side-chains only if their rewards are promising. This can be seen by shift towards lower predicted bandgaps as we traverse deeper down the tree. (d) PCA of randomly sampled and property-optimized molecules obtained from MCTS, and some popular experimental candidates. PCA is performed on Morgan fingerprints of molecules, and colors of random samples are based on Chemprop-predicted bandgaps. It can be seen that the random samples are diverse chemically and also span a range of property values, while the MCTS optimized candidates are diverse but concentrated at lower bandgap locations of the landscape. a, b, c, and d are plotted on Y6 MDP. (e) Histograms showing shifts in DFT histograms in comparison to training datasets. It can be seen that MCTS-optimized Y6 and patent molecules exhibit a large shift towards lower bandgaps compared to the patent data distribution.  }
  \label{fig:diversity}
\end{figure*}

\subsection{Generating Chemically Diverse but Optimal Candidates}
\label{diversity} We aimed to generate a number of chemically diverse, yet also high performing candidates to increase the likelihood of at least a subset of candidates being experimentally synthesizable. In addition, we have observed empirically from our previous work that the proxy reward predictor has the tendency to be adversarially attacked during training in reinforcement learning settings because of its potential unreliability in out of domain regions \cite{subramanian2023automated}. Proposing chemically diverse molecules that have high predicted rewards can allow for an increased chance that a subset of the proposals is in-domain for the reward predictor. The vanilla formulation of MCTS converges at a policy that maximizes expected cumulative reward \cite{silver2016mastering, swiechowski2023monte}, and hence does not guarantee chemical diversity. To overcome this limitation, we iteratively re-train MCTS with a time-dependent reward term that penalizes chemical similarity to high-performing molecules from previous iterations. More formally,
\begin{equation}
\label{eq:diversity}
R_i(x) = C(x) - \max_{y \in Y_{<i}} S(x, y)
\end{equation}
where $R_i$ is the reward used for training MCTS at the $i_{th}$ iteration, $C$ is the property reward function, $Y_{<i}$ is the inventory of high-performing molecules collected from iterations $i' < i$, and $S$ is a function computing similarity between molecules $x$ and $y$ (which we choose to be Tanimoto similarity). Figure \ref{fig:diversity}(a) shows the evolution of the bandgap and similarity as a function of iteration. The point displayed at each iteration corresponds to the molecule with the lowest bandgap from the last 100 timesteps of MCTS training. We can see that the constrained optimization task becomes more challenging as more molecules are appended to the inventory $Y_{<i}$, resulting in an upward trend in both blue and red curves. Figure \ref{fig:diversity}(b) zooms into a single iteration and shows the bandgap and similarity curves as a function of training step. In this case, training seems to converge after around 2,000 training steps.

\subsection{Markov Decision Process}
\label{mdp}
The Markov Decision Process (MDP) can be defined as a tuple of the state space $\mathcal{S}$, the action space $\mathcal{A}$, the reward $R$, and the transition probability $P$. We define MDPs for our two demonstrative tasks below.
\subsubsection{Y6 derivatives}
\label{y6_mdp}
We formulated this to be a highly focused and relatively small-sized chemical space as a test bed for exploring the capabilities of the approach. Building a molecule involves four hierarchies of actions to be made: 1) modification of heteroatoms and ring sizes in the Y6 \textbf{core}, 2) optional addition of \textbf{pi-bridges} to extend conjugation in the molecule, 3) addition of terminal \textbf{end-groups}, and 4) addition of \textbf{alkyl chains}. Many of the fragment choices included here were taken directly from suggestions provided in \citet{yang2021original}. We provide a more formal description of the MDP below.

Let $\mathcal{S}_t$ and $\mathcal{A}_t$ denote the state and action spaces respectively at timestep $t$. Then, $\mathcal{S}_1$ is a singleton set containing the Y6 core marked with modifiable positions shown in Figure \ref{fig:frag_decomp}(a). The first group of actions (with action spaces $\mathcal{A}_{1 \leq t \leq 4}$) are those that modify the structure of the central Y6 \textbf{core} by allowing heteroatom and ring-expansion modifications at positions marked with \texttt{pos0}, \texttt{pos1}, \texttt{pos2}, and \texttt{pos3}. 

Action spaces $\mathcal{A}_{5, 6}$ allow the optional addition of \textbf{pi-bridges} to the positions marked by $0$ to offer the flexibility to extend conjugation in the molecule, and action space $\mathcal{A}_7$ is the set of allowed \textbf{end-groups} that can be attached to $0$ (if no pi-bridge was attached) or the newly created reactive position available on the pi-bridge. Finally, $\mathcal{A}_8$ is the set of allowed \textbf{alkyl chains} that can be attached to $1$, which we restrict to just methyl groups to reduce the expense of conformer generation and optimization during DFT validation later on. This is a reasonable approximation to make given that alkyl chains are typically used to enhance solubility properties of the molecule and have a smaller influence on optoelectronic properties in isolated molecules. However, with this simplification, we disregard the more important role that alkyl chains play in optoelectronic properties when the molecules are in an aggregated state \cite{lee2021alkyl}. All fragments used in this MDP are shown in Figure \ref{fig:y6_fragments}. State space $\mathcal{S}_{t > 1}$ is the set of possible molecular structures that can be reached by applying transformations in $\mathcal{A}_{t-1}$, to state $s_{t-1}$. 

The reward $R$ is the negative of the property value (bandgap in eV) as predicted by the Chemprop \cite{heid2023chemprop} model combined with the similarity penalty term as defined in Equation \ref{eq:diversity}, and is computed once the molecular generation process has terminated. The MDP is deterministic and hence the transition probability is 1 for a particular state transition and 0 for all others at every timestep $t$, i.e., $P_t = \left( \delta_{i1} \delta_{i2} \hdots \delta_{in} \right)$ where $\delta_{ij}$ is the Kronecker delta function and $i$ is an integer ranging from $1$ to $n$ that represents the index of the chosen action. Only element $\delta_{ii}$ would be 1, and all other elements of the vector would be 0.

Figure \ref{fig:diversity}(c) and (d) showcase some features of MCTS training on the Y6 MDP. (c) demonstrates the ability of MCTS to prune the tree, with deeper actions being expanded to only if they contribute to lowering the bandgap. (d) is a principal component analysis (PCA) plot showing the partitioning of molecules in the Chemprop embedding space according to Morgan fingerprints of molecules. The MCTS optimized generations are clearly shifted towards the lower bandgap region of the embedding landscape, while also maintaining chemical diversity. They can also be seen to be red-shifted with respect to some popular experimental NFA molecules (see Figure \ref{fig:experimental_mols} for a list of the chosen experimental molecules). It is to be noted that both (c) and (d) are based on Chemprop predictions and not DFT calculations. Validation with DFT calculations is shown in later sections.

\subsubsection{Patent-Extracted Fragment Derivatives}
\label{patent_mdp}
We formulated this to be a more expansive test case with a much larger chemical space, with the aim of discovering highly red-shifted molecules with diverse chemistries. In regimes of this size, exhaustive enumeration of all achievable molecules becomes impractical.
Building a molecule in this environment involves three hierarchies of actions: 1) Choosing a \textbf{core} fragment, 2) Optional addition of \textbf{pi-bridges} to the core, and 3) Addition of terminal \textbf{end-groups/alkyl chains}. 

In this case, each reaction involves not just the choice of a fragment identity, but also the choice of reactive position to which it has to be attached. The choice of reactive position is included as stochasticity in the environment rather than an action choice in the tree. More details on this choice, and a more formal description of the MDP are provided in the following paragraphs. 

In this case, the breadth of the tree is very wide because of the large action space at each level of the tree. This makes it beneficial to trade breadth for depth of the tree. We achieve this by performing k-means clustering \cite{macqueen1967some} of fragments at each hierarchy, so that chemically similar fragments occupy the same cluster. Fragment choices are decomposed into two sequential steps of choosing a cluster, and choosing a fragment from the cluster. This approach retains the original grammar, but greatly improves efficiency of MCTS convergence. The value of $k$ was chosen arbitrarily as 100, and is a user-defined parameter that can be adjusted. Additionally, we group end-groups and alkyl chains into a single decision here unlike the Y6 case since we do not differentiate between them in the fragment decomposition algorithm described in Section \ref{decomp}.

$\mathcal{S}_1$ is a singleton set containing an empty molecule, since we build the molecule from scratch in this case. Action space $\mathcal{A}_1$ is the set of cluster indices of central \textbf{core} clusters (obtained from k-means clustering).  $\mathcal{A}_2$ is the set of fragment choices from the chosen cluster $i$, i.e., $\mathcal{C} = \{\,f \in \mathcal{F} \mid 1 \leq r(f) \leq 4; n(f) \leq 35 \,\}_i$ where $\mathcal{F}$ is the set of fragments obtained with the fragment decomposition algorithm described in Section \ref{decomp}, $r(f)$ is the number of reactive positions in fragment $f$, $n(f)$ is the number of heavy atoms (non-hydrogen atoms) in fragment $f$, and $\mathcal{X}_i$ denotes the $i$th cluster obtained from k-means clustering of set $\mathcal{X}$. We placed restrictions on the number of reactive positions and size of fragments to reduce the computational load of DFT validation.

The next group of actions contains the optional choice of \textbf{pi-bridges} to allow for extended conjugation in the molecule. This again begins with the choice of a pi-bridge cluster index, followed by the choice of fragment identity from set $P = \{\,f \in \mathcal{F} \mid r(f) = 2; n(f) \leq 10 \,\}_i$. We allow this pi-bridge sequence to be performed at most twice for a molecule (action spaces $\mathcal{A}_{3,5}$ are cluster choices and $\mathcal{A}_{4,6}$ are fragment identity choices). Finally, the last group of actions, with action spaces $A_t$ for $t \in [7, 2k+7)$, attaches end-groups to all open reactive positions on the molecule, where $k$ is the number of unique reactive positions remaining in the molecule. This attachment is iterated as long as the molecule does not exceed the maximum allowed size of 100 heavy atoms. Note that a maximum of $2k$ steps is required as we alternate between the cluster and fragment choices.
The fragment identity choice is given by set $\mathcal{E} = \{\,f \in \mathcal{F} \mid r(f) = 1; n(f) \leq 20 \,\}_i$ where $i$ is the index of the chosen cluster. State space $S_{t > 1}$ is the set of possible molecular structures that can be reached by applying transformations in $\mathcal{A}_{t-1}$, to state $s_{t-1}$. 

Similar to the Y6 case, reward $R$ is the negative of Chemprop-predicted bandgap combined with a similarity penalty as described in Equation \ref{eq:diversity}. Unlike the Y6 case however, the MDP is not deterministic here. The transitions that go from a state to a chosen fragment cluster are deterministic, but all remaining actions along the branch have some randomness associated with them. Once an action $a_t$ corresponding to the identity of pi-bridge or end-group is chosen, the reactive position on the current state $s_t$ is sampled from a uniform distribution, and the next state $s_{t+1}$ is reached by reacting the fragment at the sampled reactive position on the current state. 
The transition probability is therefore defined as
\begin{equation*}
    P_t = \begin{cases}
        {\left( \delta_{i1} \, \delta_{i2} \, \cdots \, \delta_{in} \right)} & {t \text{ is odd}} \\
        {\left( \frac{1}{k_t} \, \frac{1}{k_t} \, \cdots \, \frac{1}{k_t} \right)} & {t \text{ is even}} \\
    \end{cases}
\end{equation*}

$k_t$ is the number of unique reactive positions available on state $s_t$. We chose to include this stochasticity in reactive position choice for two reasons: 
1) The identity and nature of a chemical fragment typically has a higher influence on the optical properties of a molecule, than the reactive position choice does. Hence removing the reactive position choice from the action space allowed us to limit the depth of the tree and improve computational efficiency during training.
2) We wanted MCTS to converge at solutions that identify optimal fragment choices but are relatively robust to choice of reactive positions. While the precursors are determined by the fragments that are present in the molecule, there is more flexibility in the reactive position choice, and therefore during synthetic reaction planning.

\subsection{Training Reward Predictor}

\subsubsection{TD-DFT Calculations} \label{tddft} TD-DFT calculations were used as ``ground-truth'' labels to train our Chemprop reward predictor, as well as to validate the bandgaps of generated candidates.
Calculations were performed with the same pipeline that we used in \citet{subramanian2023automated}. 
RDKit ETKDG \cite{landrum2013rdkit} approach was used to generate initial conformations, with at least 1,500 attempts. The conformers were ranked according to their MMFF94 energies, and the 20 with the lowest energies were chosen \cite{riniker2015better}. An initial geometry optimization was performed with semi-empirical tight-binding density functional theory (GFN2-xTB) \cite{Bannwarth2019} in ORCA 4.2.0 \cite{Neese2020}. Subsequent geometry optimizations were performed on the lowest energy xTB conformers at the BP86\cite{Becke1988}-D3\cite{Grimme2011}/def2-SVP\cite{Weigend2005} level of theory. Finally, single-point TD-DFT calculations were performed at the $\omega$B97X-D3\cite{Chai2009}/def2-SVPD level of theory using the Tamm--Dancoff approximation (TDA) \cite{hirata1999time} on the optimized geometries. Bandgaps used for reward training were the lowest-energy singlet vertical excitation energies from the TD-DFT calculations.

\begin{figure*}[ht]
  \centering
  \includegraphics[width=\textwidth]{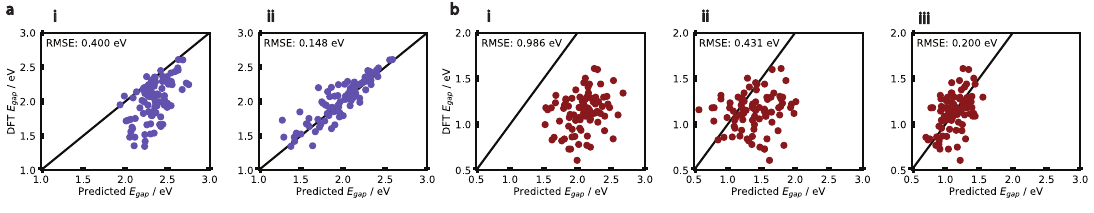}
  \caption{\textbf{Active Learning to improve reward prediction}  Scatter plots showing improvement in fit with AL iterations for (a) Y6 derivatives (b) Patent-extracted fragment derivatives. The test data in (a) and (b) are the final sets of 100 molecules generated after all AL iterations have been completed with the Y6 and patent-fragment MDPs respectively. In (a) and (b), the plot i corresponds to the model pre-trained on just patent-mined dataset, and plot ii corresponds to model trained on patent dataset + random rollouts from MCTS. Plot iii in (b) corresponds to model trained on patent dataset + random rollouts + diversity \& EI acquisition samples. More details are given in Section \ref{AL}. }
  \label{fig:AL}
\end{figure*}

\subsubsection{Active Learning for Data Collection}
\label{AL}
To ensure that the reward predictor is accurate in the generation domain of MCTS, we performed iterations of active learning (AL) with a combination of acquisition functions to collect training data. Figure \ref{fig:AL} shows the improvement in reward predictor with each iteration of AL. Chemprop \cite{heid2023chemprop} was our architecture of choice for the reward predictor. It was first pre-trained on a dataset containing 5,568 datapoints obtained from patents and labeled with TD-DFT bandgaps which we created in \citet{subramanian2023automated}. The second round of training was performed on random rollouts generated from MCTS. These two rounds were sufficient to get accurate (we chose the stopping criterion to be an RMSE of 0.2 eV) performance on the focused Y6 chemical space, but the patent-extracted fragment space required an additional round of training. This was done on a combination of datapoints obtained from 1) running the iterative procedure described in Section \ref{diversity} and choosing the best candidate from the last 100 training steps of each iteration, and 2) choosing 100 molecules with the highest value of expected improvement (EI) from 10,000 randomly sampled candidates from MCTS.
\begin{equation*}
\label{eq:EI}
EI(x) = \left( \mu - f(x^\star) \right) \Phi \left( \frac{\mu - f(x^\star)}{\sigma} \right) + \sigma \phi \left( \frac{\mu - f(x^\star)}{\sigma} \right)
\end{equation*}
where $\mu$ is the negative of the Chemprop-predicted bandgap, and $\sigma$ is the standard deviation of predictions from the model ensemble, which is a measure of model uncertainty. $\Phi$ and $\phi$ are the cumulative distribution function (CDF) and probability density function (PDF) of a normal distribution, respectively. $f(x^\star)$ is the highest value of predicted reward from the current batch of candidates. While acquisition method 1 obtains a diverse set of most optimal candidates, method 2 obtains a set of optimal candidates as well as suboptimal but high-uncertainty candidates. Together, these two acquisition methods provide training data in explorative and exploitative regimes.


\begin{figure*}[ht]
  \centering
  \includegraphics[width=\textwidth]{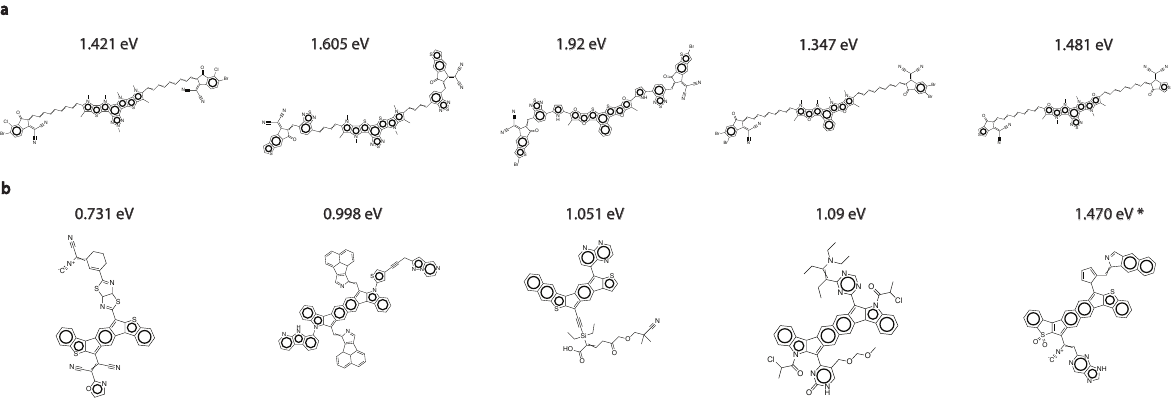}
  \caption{\textbf{Final generated candidates and their TD-DFT bandgaps} K-means clustering was performed on final 100 molecules from each category into 5 clusters (based on Morgan fingerprints). The molecules shown are the lowest-bandgap molecules chosen from each cluster. (a) Y6-derivatives MDP, (b) Patent-extracted fragments MDP. Bandgap of the last molecule is denoted with asterisk (*) because the geometry had to be fixed before the TD-DFT calculation was performed. More details are provided in Section \ref{fixes} of SI.}
  \label{fig:molecules}
\end{figure*}

\section{Results and Discussion}
\label{results}
We validated the final 100 diverse candidates obtained from each of Y6 and patent MDPs with TD-DFT bandgap calculations. We show histograms of these bandgaps in Figure \ref{fig:diversity} (e) in comparison to the original patent dataset, and random rollouts sampled from MCTS. It can be seen that the MCTS optimized patent and Y6 candidates are strongly red-shifted with respect to random rollouts as well as the patent distribution.

On the final 100 structures from each MDP, we performed k-means clustering with $k=5$ based on Morgan fingerprints of molecules to obtain the 5 most chemically diverse clusters of molecules. We then chose the molecule with the lowest TD-DFT bandgap from each cluster. These structures along with their bandgaps are displayed in Figure \ref{fig:molecules}. It can be seen that these molecules display structural features that are to be expected from low-bandgap molecules, such as highly conjugated rings. Furthermore, the predicted bandgaps all fall in the NIR region of the spectrum. 

In the case of Y6 derivatives, it is interesting to see a higher tendency of MCTS to generate long polyene bridges between cores and end-groups (see the first, fourth, and fifth molecules in subfigure (a)). Clearly, it has exploited the correlation between degree of conjugation and lowering of bandgap. However, the presence of conical intersections (CI) in long linear polyenes is well known \cite{fuss2000twin,nenov2012conical} suggesting a possibility for these MCTS-predicted molecules to be photochemically inactive. Linear response TD-DFT has been seen to be unsuitable for accurately predicting the presence of these points, paving way for the development of more sophisticated techniques such as spin-flip TD-DFT (SF-TD-DFT) \cite{levine2006conical,minezawa2009optimizing}. This is hence a drawback with the complete reliance of MCTS on the accuracy of TD-DFT predictions, but nevertheless showcases the ability of the algorithm to reliably optimize the chosen reward.

Design strategies such as spatially separating donating and accepting fragments for a ``push-pull'' effect are also known to have a lowering effect on bandgaps \cite{pirotte2018true}. Among the generated patent-derivatives, this design pattern can be seen in the first and fourth molecules of subfigure (b) which contain electron-rich cores and electron-withdrawing end-groups. We additionally observe the presence of highly reactive sites in some of these structures (such as hydrogen atoms on aromatic amines of second, fourth, and fifth molecules, and chlorine atoms on the fourth molecule). We replace the reactive positions with methyl groups and perform TD-DFT calculations on the ``corrected'' structures to verify that optical properties do not drastically change even after the substitution. These results can be found in Section \ref{fixes} of SI.

Finally, the TD-DFT bandgap calculations that we report work under the assumption of 0 K temperature. Furthermore, a more realistic device-like environment would contain thin-films containing several molecules packed together. We perform coupled MD/TD-DFT calculations on the final 10 molecules to better represent finite temperature and molecular aggregation effects. We report absorption spectra calculated with these methods, and more details about this methodology in section \ref{aggregation} of the SI.

\section{Conclusions}
In this paper, we developed an approach based on the Monte Carlo Tree Search algorithm to generate optimal low-bandgap molecules while also maintaining a degree of synthetic accessibility. We achieve this by utilizing structural priors from a domain-focused patent-mined dataset of organic electronics molecules through a symmetry-aware fragment decomposition algorithm and a fragment-constrained MCTS generator. Our generated candidates retain symmetry constraints from the patent dataset while also exhibiting red-shifted absorption. As future extensions to this work, the optimization objective can be augmented to include solubility (along with the inclusion of alkyl chain exploration into the MDP) and oscillator strengths. Finally, while our approach for diverse candidate generation is iterative, non-iterative objectives as used in GFlowNet \cite{bengio2023gflownet} could be more efficient alternatives for the future.

\section{Data and Code Availability}
All data and code accompanying this article, are available at \url{https://github.com/learningmatter-mit/symmetry-mcts}.

\section{Acknowledgements}
A. S. was supported by funding from Sumitomo Chemical. K. P. G. was supported by the National Science Foundation Graduate Research Fellowship Program under Grant No. 1745302. A. P. S. was supported by the Department of Defense (DoD) through the National Defense Science and Engineering Graduate (NDSEG) Fellowship Program. We acknowledge the MIT Engaging cluster and MIT Lincoln Laboratory Supercloud cluster \cite{reuther2018interactive} at the Massachusetts Green High Performance Computing Center (MGHPCC) for providing high-performance computing resources to run our TD-DFT calculations and train our deep learning models. We also thank Prof.~Aristide Gumyusenge, Saumya Thakur, Jurgis Ruza, Soojung Yang, Ajay Subramanian, Sulin Liu, and all members of the RGB group for useful discussions throughout.

\bibliography{main}
\bibliographystyle{icml2021}

\newpage
\appendix
\label{appendix}
\onecolumn

\section{Acquisition Function for AL}
Expected improvement is an bayesian acquisition function that builds on top of probability of improvement. While a more detailed discussion can be seen in \citet{kamperis2021acquisition}, we reproduce a brief excerpt of the derivation below.
\subsection{Probability of Improvement (PI)}
Given a function $f$, our task is to estimate the probability that $f(x) > f(x^\star)$ where $x^\star$ is a previous optimum. If improvement is defined as, 
\begin{equation*}
\label{eq:EI}
I(x) = max(0, f(x) - f(x^\star)),
\end{equation*}
If $f(x)$ is treated as a random variable following the gaussian distribution $\mathcal{N}(\mu, \sigma^2)$, then using the reparameterization trick, we can rewrite $I(x)$ as
\begin{equation*}
\label{eq:EI}
I(x) = max(0, \mu(x) + \sigma(x)z - f(x^\star)), z \sim \mathcal{N}(0, 1)
\end{equation*}
Then the probability of improvement
\begin{equation*}
\label{eq:EI}
PI(x) = Pr(I(x) > 0) = \Phi(\frac{\mu(x) - f(x^\star)}{\sigma(x)}),
\end{equation*}
where $\Phi(z) = CDF(z)$.

\subsection{Expected Improvement (EI)}
EI, unlike PI, is the expected value of improvement $I(x)$, which gives us an estimate of the magnitude of improvement rather than just the probability of improvement.

It is defined as 
\begin{equation*}
\label{eq:EI}
EI(x) = \int_{-\infty}^{\infty}I(x)\varphi(z)dz = \int_{-\infty}^{\infty}max(0, f(x) - f(x^\star))\varphi(z)dz,
\end{equation*}
where $\varphi(z) = \frac{1}{\sqrt{2\pi}}exp(-z^2/2)$.
The integral can be split as 
\begin{equation*}
\label{eq:EI}
EI(x) = \int_{-\infty}^{z_0}I(x)\varphi(z)dz + \int_{z_0}^{\infty}I(x)\varphi(z)dz
\end{equation*}
The first term is zero since $I(x) = 0$.

\begin{equation*}
\begin{split}
EI(x) &= \int_{z_0}^{\infty}(\mu + \sigma z - f(x^\star))\varphi(z)dz \\
&= \int_{z_0}^{\infty} \left(\mu - f(x^\star) \right)\varphi(z)\mathop{\mathrm{d}z} +
\int_{z_0}^{\infty} \sigma z \frac{1}{\sqrt{2\pi}}e^{-z^2/2}\mathop{\mathrm{d}z} \\
&=\left(\mu- f(x^\star)\right) + \sigma \varphi(z_0) \\
&=\left(\mu- f(x^\star)\right) \Phi\left(\frac{\mu-f(x^\star)}{\sigma}\right) + \sigma \varphi\left(\frac{\mu - f(x^\star)}{\sigma}\right),
\end{split}
\end{equation*}

where $\Phi$ and $\varphi$ are the cumulative distribution function (CDF) and probability density function (PDF) of a normal distribution, respectively.

\begin{figure*}[ht]
  \centering
  \includegraphics[width=0.8\textwidth]{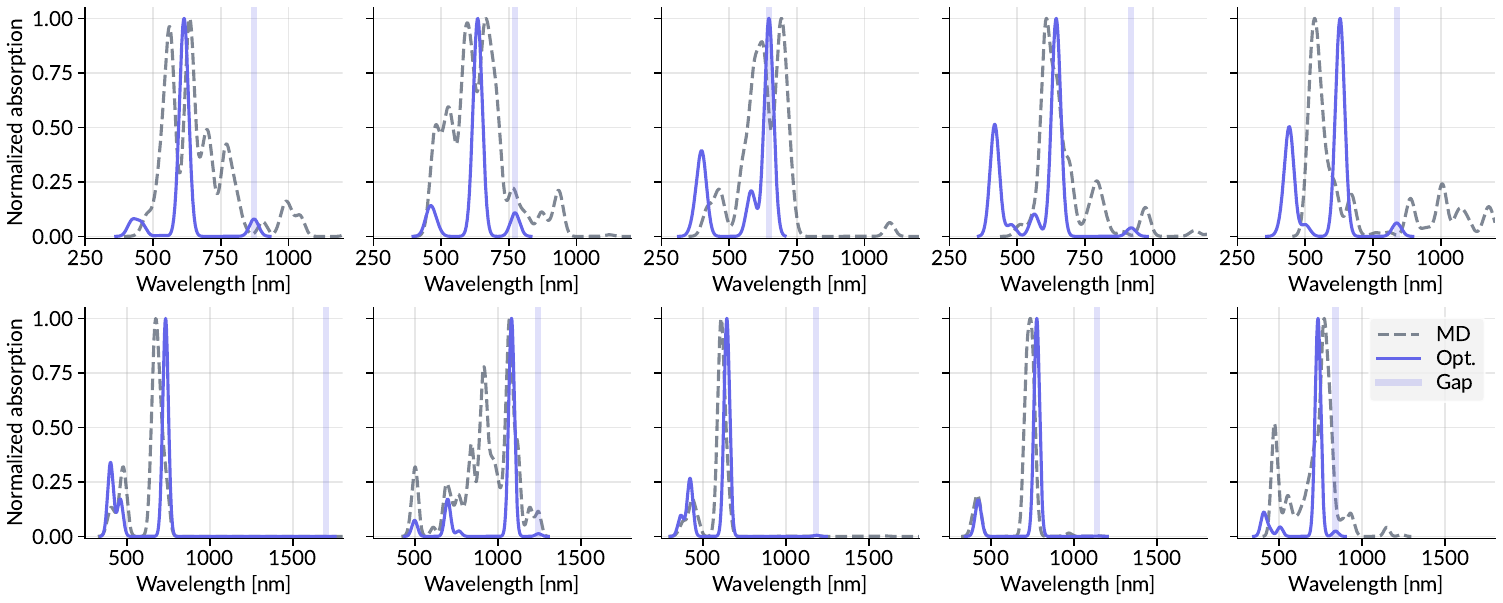}
  \caption{\textbf{Computed absorption spectra for final candidate molecules.} The spectra for molecules are ordered as shown in Figure 4 in the main text, with Y6 derivatives in the upper row and patent-extracted fragment designs in the lower row. Solid lines represent single-point TD-DFT calculations at the optimized geometry, while dotted lines represent statistically averaged TD-DFT spectra from molecular clusters sampled from MD simulations. The translucent vertical lines indicate the band gaps from the single-point calculations. The spectra are normalized so that the maximum absorption corresponds to 1.0.}
  \label{fig:absorption_spectra}
\end{figure*}

\section{Finite Temperature and Aggregation Effects with Coupled MD/TD-DFT Calculations}
\label{aggregation}

\subsection{Computational Methods}
We combined MD simulations with TD-DFT calculations to approximate the effects of finite temperature and molecular aggregation on absorption spectra. Following \citet{kupgan2021molecular}, we simulated an amorphous morphology for the molecules. Starting with the optimized geometries presented in the main text, we used PACKMOL \cite{martinez2009packmol} to pack 200 structures in a cubic box at a low density ($\sim$ 0.1 g/cm$^3$). The OPLS-AA force field \cite{robertson2015improved} was used via the LigParGen server \cite{dodda2017ligpargen}. The system was equilibrated for 30 ns at 650 K, cooled to room temperature (300 K) at 10 K/ns, and subjected to a 30 ns production run at 300 K. All MD simulations were performed under the NPT ensemble at 1 atm, using GROMACS 2023 \cite{abraham2015gromacs}. From the last 20 ns of the production run, we randomly selected a molecule and included all neighboring molecules within a 5 \AA{} radius. CHELPG charges \cite{breneman1990determining} were computed for each neighboring molecule, and TD-DFT calculations were performed for the selected molecule with the neighboring point charges. For electronic structure calculations, we used the same settings as in the main text. The absorption spectra were averaged over 10 clusters to obtain a statistical estimate.

Two of the patent-derived molecules (the first and fifth molecules in Figure 4 of the main text) contain an isocyanide moiety, which could not be parametrized correctly using the LigParGen server. Therefore, for these two molecules, we substituted the isocyanide (--NC) functional group with cyanide (--CN) for this analysis. This substitution does not impact the band gap calculations significantly, with differences in bandgaps for the optimized geometries of only 2 and 6 meV, respectively, indicating negligible effects on the result analysis.

\subsection{Absorption Spectra of Final Candidates}

The absorption spectra for the designed molecules are presented in Figure~\ref{fig:absorption_spectra}, with the Y6 derivatives in the upper row and patent derivatives in the lower row. The vertical lines indicate the TD-DFT band gap for the optimized geometries, as reported in Figure 4 of the main text. The MD/TD-DFT pipeline results for the primary absorption peak locations are consistent with those from static (optimized) TD-DFT calculations, though the MD/TD-DFT spectra exhibit broadening, likely due to thermally accessible conformers and neighboring molecules. Note that the reddest absorption peak from TD-DFT corresponding to the band gap might not accurately reflect the absorption spectra when its oscillator strength is very low. This limitation, inherent to the design objective focused solely on the band gap, indicates the need to consider transition probability in future designs. Nonetheless, the band gap serves as a robust design objective, and this result demonstrates that our MCTS design pipeline effectively produces low-bandgap molecules.

\section{Preprocessing of Patent-Mined Fragments}
After fragment decomposition was performed, we performed some minimal preprocessing steps to ensure that the fragments are representative of the chemical space we want to explore. We first performed an RDKit filter, to ensure that only SMILES strings representing valid molecules are retained, followed by an element filter, which retained only fragments containing the following elements: \texttt{C}, \texttt{O}, \texttt{N}, \texttt{H}, \texttt{Cl}, \texttt{Br}, \texttt{S}, \texttt{F}, \texttt{I}, and \texttt{Si}. We finally removed fragments that do not contain any aromatic atoms.

\begin{figure*}[ht]
  \centering
  \includegraphics[width=0.8\textwidth]{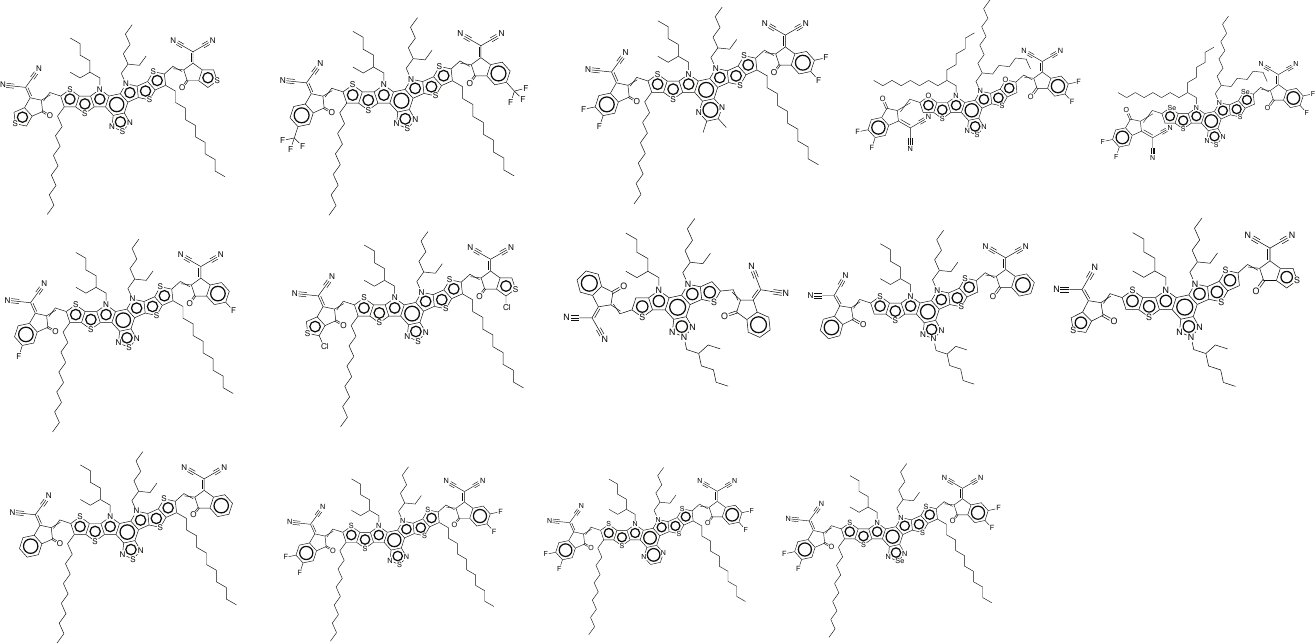}
  \caption{\textbf{Some popular experimentally used acceptor molecules}. These were chosen from \citet{lu2021recent}. }
  \label{fig:experimental_mols}
\end{figure*}

\section{Correlation Between DFT and Experimental Bandgaps}
\begin{figure}[ht]
  \centering
  \includegraphics[width=0.4\columnwidth]{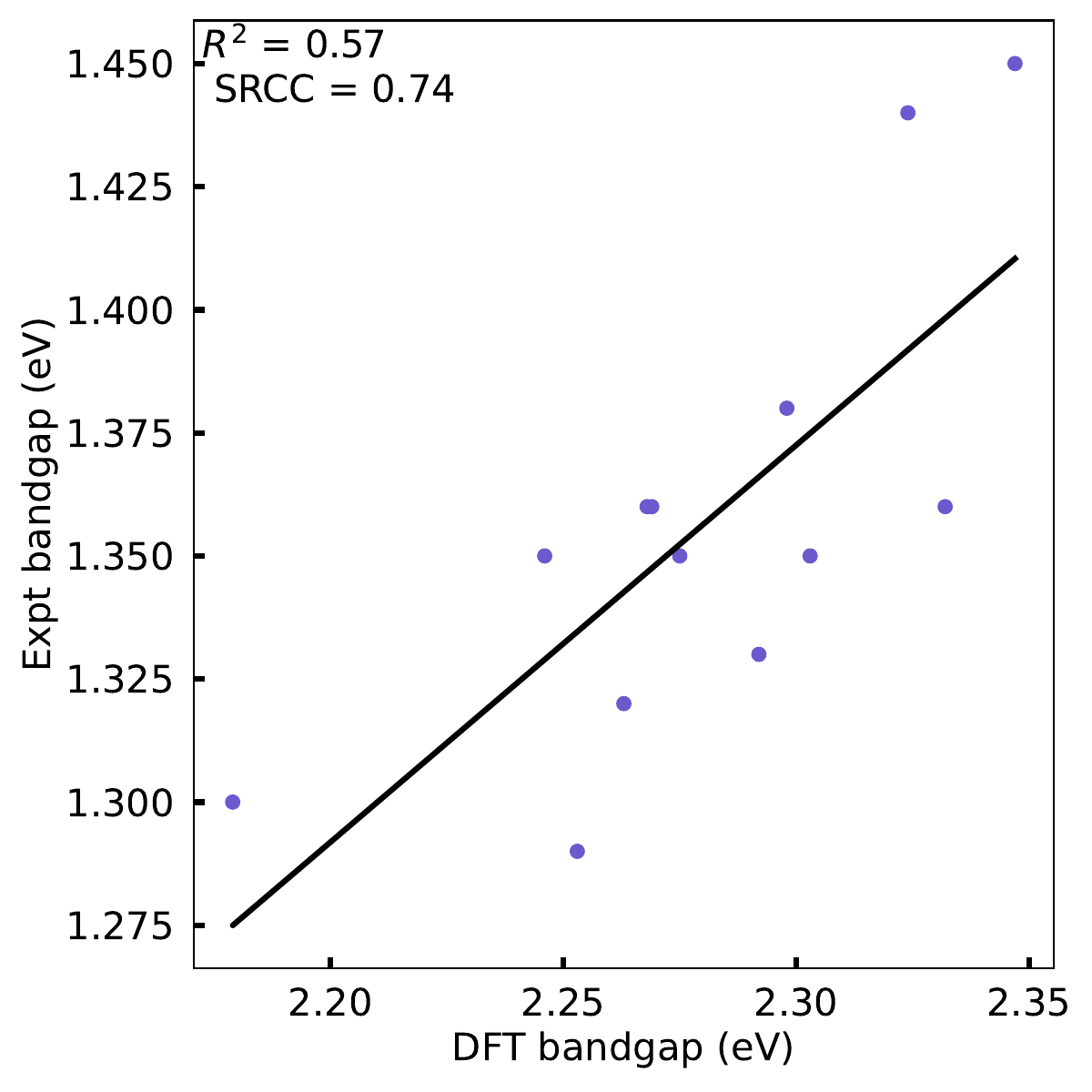}
  \caption{\textbf{Correlation between TD-DFT and experimental bandgaps.} $R^2$ score and Spearman rank correlation coefficient are shown on the top left.}
  \label{fig:dft_expt}
\end{figure}

We performed TD-DFT calculations on a subset of experimental molecules from the list shown in Figure \ref{fig:experimental_mols} for which we had access to experimentally measured bandgaps. The correlation plot between experimental and DFT values are shown in Figure \ref{fig:dft_expt}. We see $R^2$ value of 0.57, and a Spearman rank correlation coefficient (SRCC) of 0.74. While the magnitudes are not well calibrated, the trends of experiments are well-captured by TD-DFT, making minimization a suitable objective for MCTS as opposed to optimization for targeted bandgap values. As a future extension of this work, it could be possible to fit a calibration function between TD-DFT and experimental bandgaps and use the calibrated rewards for training. While it could be somewhat reasonable to extrapolate such calibrations for the Y6 MDP, it might be more challenging to obtain representative experimental examples for the patents MDP on which a calibration fit can be performed. 

The mismatch in magnitudes between TD-DFT and experiments can be caused by several factors including (but not limited to) 1) finite temperature effects, 2) molecular aggregation effects from experimental values being measured on thin films, 3) inherent error in TD-DFT functional. 

\section{Structural Modifications}
\label{fixes}
For the fifth molecule shown in Figure 4 of the main text, we identified that errors in atomic connectivity were introduced during the xTB optimization step. We fixed the geometry by bypassing the xTB optimization step altogether. We instead directly performed BP86 DFT optimization on the five lowest energy conformers obtained from RDKit ETKDG algorithm, and performed the TD-DFT calculation on the lowest energy BP86 conformer. 

As mentioned in Section 3 of the main text, we performed TD-DFT calculations on methyl-substituted versions of reactive molecules to confirm that the bandgap does not change significantly after substitution. Bandgaps of second, fourth, and fifth molecules changed very minimally from 0.998, 1.09, and 1.470, to 0.978, 1.049, and 1.470 eV respectively.

\section{Reactive Positions in Practice}
\label{pos_practice}
While we explain the algorithm in Section 2.1.1 of the main text using IDs belonging to $[0,N)$, in the implementation, we use Helium isotopes with IDs ranging from  $[100,100+N)$ where $N$ is the number of fragments in the vocabulary, to preserve reactive position identities in the RDKit \texttt{Mol} object.
The range was chosen arbitrarily but to be large enough so that there are no clashes in atomic mass with other elements in the dataset. 

\section{Fragments in Y6 MDP}
\begin{figure*}[ht]
  \centering
  \includegraphics[width=0.8\textwidth]{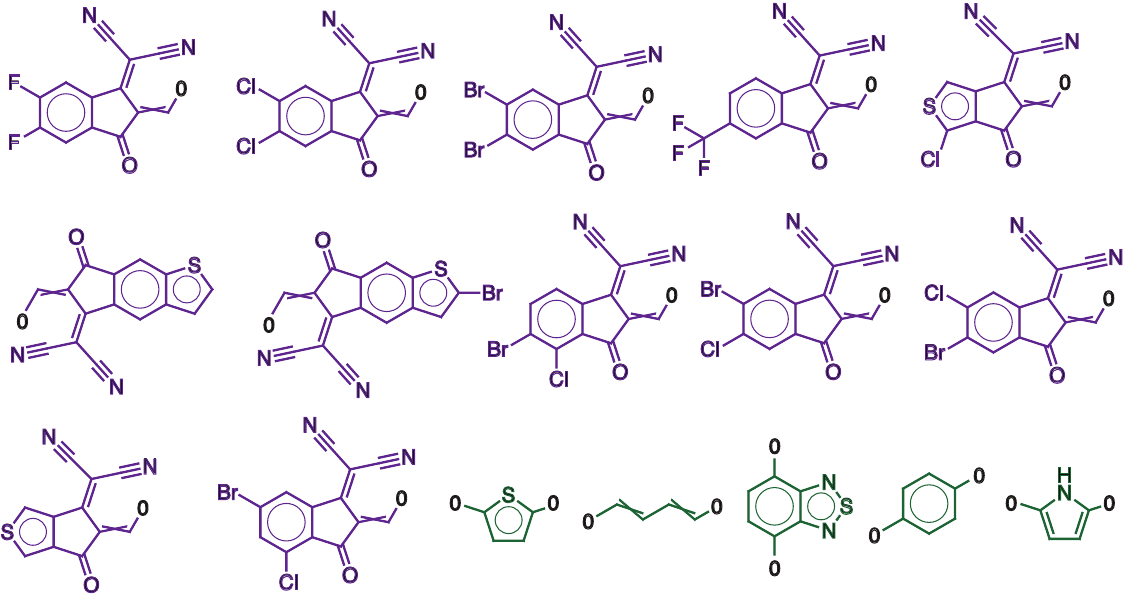}
  \caption{\textbf{Fragments used in Y6 MDP} end-groups (purple) and pi-bridges (green). Core-modification actions are implemented as string replacements, the list of which can be found in the GitHub repository. We also omit illustrating alkyl chain actions here since methyl group is the only choice explored.}
  \label{fig:y6_fragments}
\end{figure*}

We illustrate the fragments along with their reactive positions in Figure \ref{fig:y6_fragments}.

\end{document}